\documentclass[letterpaper]{article} 
\usepackage[submission]{aaai23}  
\usepackage{times}  
\usepackage{helvet}  
\usepackage{courier}  
\usepackage[hyphens]{url}  
\usepackage{graphicx} 
\usepackage{natbib}  
\usepackage{caption} 
\usepackage[ruled, vlined, linesnumbered]{algorithm2e}
\usepackage{comment}
\usepackage{cite}
\usepackage{subfig}

\usepackage{mwe}
\usepackage{subfig}
\usepackage{amsmath,amssymb,amsfonts,amsthm}
\usepackage{multirow}
\usepackage{xcolor}
\usepackage{array}
\usepackage{makecell}

\title{Predicting Customer Goals in Financial Institution Services: \\ A Data-Driven LSTM Approach}

\author{
Andrew Estornell$^*$,\textsuperscript{\rm 1}
Stylianos Loukas Vasileiou\footnote{Equal contribution.},\textsuperscript{\rm 1}
William Yeoh,\textsuperscript{\rm 1}
Daniel Borrajo,\textsuperscript{\rm 2}
Rui Silva\textsuperscript{\rm 2}
}
\affiliations {
\textsuperscript{\rm 1} Washington University in St. Louis\\
\{aestornell,vstylianos, wyeoh\}@wustl.edu \\ [2pt]
\textsuperscript{\rm 2} J.P. Morgan AI Research\\
\{daniel.borrajo,rui.silva\}@jpmchase.com 
}

\begin{document}

\maketitle

\begin{abstract}
In today's competitive financial landscape, understanding and anticipating customer goals is crucial for institutions to deliver a personalized and optimized user experience. This has given rise to the problem of accurately predicting customer goals and actions. Focusing on that problem, we use historical customer traces generated by a realistic simulator and present two simple models for predicting customer goals and future actions -- an LSTM model and an LSTM model enhanced with state-space graph embeddings. Our results demonstrate the effectiveness of these models when it comes to predicting customer goals and actions.
\end{abstract}

\section{Introduction}
The financial industry has experienced significant transformation in recent years, driven by rapid technological advancements, evolving customer expectations, and increased competition. As customers demand more personalized and convenient services, financial institutions are under pressure to develop a deeper understanding of their clients' needs and preferences. This has led to a growing interest in leveraging data-driven approaches to gain insights into customer behavior and predict future actions. Accurate goal prediction can help financial institutions provide targeted incentives, improve customer satisfaction, and ultimately foster loyalty and retention in an increasingly competitive landscape.

Planning-based approaches have been widely used for goal prediction, as they focus on modeling an agent's decision-making process and finding optimal sequences of actions to achieve specific objectives \cite{ramirez2010probabilistic,sohrabi2016plan,DBLP:journals/jair/KerenGK19,wayllace2016goal,veredheuristic}. For example, \citet{ramirez2010probabilistic} propose a probabilistic plan recognition approach that utilizes classical planners to predict goals based on observed actions.~\citet{DBLP:journals/jair/KerenGK19} introduced the concept of goal recognition design, which aims to optimize the planning domain to facilitate the goal prediction process.

In this context, the work of \citet{borrajo2020domain} presents an approach to address the challenges associated with predicting goals in complex real-world domains. In particular, they introduced a domain-independent simulator for generating synthetic customer behavior datasets, which can be used to model and analyze customer interactions with a financial institution, such as ATM or mobile app usage. By leveraging this simulator, they are able to create datasets that accurately represent the intricacies and dynamics of customer traces in a banking environment, providing a valuable foundation for the development and evaluation of goal prediction models.

Inspired by the work of \citet{borrajo2020domain}, we build upon their domain-independent simulator to generate synthetic customer behavior datasets. We then use this data to train two models -- an LSTM model and an LSTM model enhanced with state-space graph embeddings. The LSTM-based models capitalize on the sequential nature of customer traces, capturing the intricate patterns present in customer interactions over time. By incorporating state-space graph embeddings into the LSTM model, we further enrich the model's understanding of the relationships and dependencies among various features within the dataset, which may lead to improved performance. This combination of LSTM models and state graph embeddings offers a more scalable and efficient solution in predicting customer goals and actions, while maintaining a high level of accuracy and robustness in the face of real-world complexities. Our results demonstrate the effectiveness of LSTM models and state graph embeddings in addressing the challenges faced by financial institutions when it comes to predicting customer goals and actions.

\section{Related Work}

Goal prediction (known as goal recognition in the planning literature) refers to the process of identifying an agent's objective among several possibilities, based on the agent's behavior, a model of the environment, and a sequence of observations. The first approach that explicitly addresses goal prediction is that of \citet{inverseplanning}, who framed the problem as ``inverse'' planning. \citet{ramirez2010probabilistic} followed up on the work by proposing a generative approach that uses classical planning algorithms for inferring probability distributions over a set of possible goals given some observations. Recent research has extended the original problem formulation of recognizing plans in various domain models, such as continuous domain models and epistemic planning problems \cite{veredheuristic,sohrabi2016plan,kaminka2018plan,shvo2020epistemic}. In similar spirit, \citet{DBLP:journals/jair/KerenGK19} provided an alternate view of goal recognition that focuses on modifying the domain model such that the goal recognition can be achieved with as few observations as possible. Extensions of the goal recognition design problem have been developed over various types of domain models and settings \cite{shvo2020active,wayllace2016goal,wayllace:20,wayllace22}.

However, planning-based approaches often struggle to scale with the size and complexity of real-world domains, due to factors such as large state spaces, partial observability, and dynamic environments. In contrast, learning-based approaches may perform better under these circumstances, especially when real-time or near-real-time predictions are required. Most notably, \citet{borrajo2020goal} illustrated the trade-offs between planning-based and learning-based approaches for goal recognition in different (planning) domains. Their results highlight that planning-based approaches perform better when there is a partial order of actions in plans, whereas learning techniques, such as LSTMs, can better capture the relationship between actions and goals if there is a relation not directly associated with goal achievement. While in this paper we also use a learning-based approach (e.g., LSTMs) for goal prediction, note that we focus on the specific task of predicting customer goals and actions in the context of financial institutions. Finally, there has been an interest in integrating planning-based approach with data-driven approaches. For example, \citet{wilken2023investigating} proposed a hybrid method that captures statistical relations between certain states of the environment and goals learned from past observations. 

While in this paper we focus on goal prediction problems, i.e., predicting user goals and actions, we explore data-driven methodologies and focus on a problem related to financial industry. In that context, a number of works that tackle related problems have been investigated. \citet{baeza2015predicting} presented a method for predicting the next mobile app a user is going to open based on their usage history and the ``wisdom'' of the crowd, while several works have addressed the problem of customer churn in banking with various machine learning techniques \cite{yaghini2011prediction,bilal2016predicting,rahman2020machine}. While these works and our work in this paper address challenges in the financial industry, the specific goals and techniques employed are different, i.e., we employ LSTM models with graph embeddings aimed at understanding customer behavior and preferences.

\section{Methodology}
In this section, we describe the dataset, data preparation, and models used for predicting customer goals and actions. In particular, we employ two models -- an LSTM model exploiting a bag-of-words representation, and an LSTM model enhanced with a state graph embedding. Our approach is underpinned by a comprehensive semi-synthetic dataset, capturing intricate customer interactions with diverse banking interfaces.

\begin{table}[t]
    \centering
    \begin{tabular}{|c|c|}
    \hline
       \textbf{Date and Time} & \textbf{Event} \\ \hline\hline
    2022-02-28 18:06:08	& \makecell{mobile: login} \\\hline
    2022-02-28 18:10:24	& \makecell{mobile: enter menu settings} \\\hline
    2022-02-28 18:10:24	& \makecell{mobile: enter menu \\ profile-maintenance} \\\hline
    2022-02-28 18:14:40	& \makecell{mobile: change information \\ on demographic} \\\hline
    2022-02-28 18:18:56	& \makecell{mobile: log-off} \\\hline\hline
    2022-04-30 07:28:32	& \makecell{web: login} \\\hline
    2022-04-30 07:28:32	& \makecell{web: enter menu \\ credit-card} \\\hline
    2022-04-30 07:28:32	& \makecell{web: get information on \\credit-card-transaction-history} \\\hline
    2022-04-30 07:28:32	& \makecell{web: exit menu \\root-section} 	\\\hline
    2022-04-30 07:32:48	& \makecell{web: enter menu settings} \\\hline
    2022-04-30 07:37:04	& \makecell{web: enter menu \\alerts-maintenance} \\\hline
    2022-04-30 07:41:20	& \makecell{web: change information\\ on alerts-definition} \\\hline
    2022-04-30 07:45:36	& \makecell{web: log-off} \\\hline
    \end{tabular}
     \caption{Example trajectory of events for an agent of \textit{medium income}, \textit{low fail behavior}, and \textit{digital interface}.}
    \label{tab:example_trajectory}
\end{table}

\subsection{Customer Behavior Data}

We utilize a semi-synthetic dataset of customer behavior generated via a domain-independent simulator proposed by \citet{borrajo2020domain}. 
The dataset consists of interactions between customers and a multitude of bank interfaces: bank-website, mobile app, teller counter, or ATM. 
Each interface offers customers a (potentially unique) set of actions such as making payments, checking rewards program information, modifying personal information, etc.
A set of example actions are provided in Table~\ref{tab:example_trajectory}, where we see a customer login to the mobile app and modify their personal information; the same customer then logs into the website two months later and checks their credit-card transaction history as well as modifying their alter settings.
Approximately $300$ actions are recorded per-customer.

\paragraph{Customer Goals:} In addition to customer actions, a diverse set of goals can be induced from the dataset, such as:
\begin{itemize}
\item \textit{check information}: The customer wants to get some information about their account (e.g., balance).
\item \textit{change information}: The customer wants to change the value of some data point (e.g., address).
\item \textit{operational goals}:  The customer wants to perform some banking operations, such as deposit-cash, withdrawal, exchange, deposit-check, pay-bill, make-payment, and so on. 
\end{itemize}
These goals correspond to the customers purpose for interacting with any of the bank interfaces. 
Customers can possess multiple goals; for example, the customer in Table~\ref{tab:example_trajectory} seeks to both check and change information while using the website (second interaction in the table).

\paragraph{Customer Types:} The dataset also features labels for the type of customer, characterized by three attributes: 
\begin{itemize}
    \item \textit{Income}:         income level of the customer (high, medium, low, standard).
    \item \textit{Fail behavior}:  how frequently the customer's action correspond to errors (rarely, often, no-failure).
    \item \textit{Digital behavior}: customers preferred interface type (traditional, digital, mixed).
\end{itemize}

\noindent These attributes represent the probabilities of using different channels (web, mobile app, ATM, banker, and teller), the failure rate of operations, and types of goals. For instance, a student might be categorized as low-rarely-digital, while a medium-class worker could be medium-rarely-mixed.

\subsection{Data Preparation}
Next we discuss our approach to data preprocessing. 
The predictive features $\mathbf{X}$ are the customer's historical actions (e.g. Table~\ref{tab:example_trajectory}). 
The target features $y$ are the customer's future actions, the customer's goal, or the customer's type.

\paragraph{Handcrafted Features:}
To increase the predictive efficacy of the data we first introduce several handcrafted features, and then discuss how both the handcrafted, and original features, are represented.
The predictive features $\mathbf{X}$ are provided as Time (an integer) and Action (a string). 
We break the Event string into more meaningful features which are listed in Table~\ref{tab:handcrafted_features}. The Event string contains an indicator of which interface the customer is interacting with (web, mobile, teller, ATM), which we refer to as the \emph{primary location} feature.

Moreover, the Event string contains information indicating when the customer navigates through the interface, e.g., in Table~\ref{tab:example_trajectory} the customer has Event string ``web: enter menu credit-card", which indicates that the customer has left the web home-page (which they arrived at when logging in) and navigated to a menu of credit-card options. We refer to these features as \emph{secondary location} features.

Defining secondary locations allows the model to more effectively learn which actions are available to customers at any given time, e.g., customers cannot modify their credit-card information without having first entered the credit-card menu.
The combination of the primary and secondary location features can be interpreted as the customer's current state.

In addition to the state features, we also define corresponding \emph{action} features that fall into three categories: \emph{transitioning-actions} that indicate the customer changing primary (or secondary) locations (e.g., ``web: enter menu credit-card"); \emph{information-gaining-actions} that indicate the customer obtaining new information (e.g. ``web: get information on credit-card-transaction-history"); and \emph{modification-actions} that indicate the customer modifying their information (e.g. ``web: change information on alerts-definition"). 
Full details on the state-action features are provided in Table~\ref{tab:handcrafted_features}.
The predictive features $\mathbf{X}$ are thus represented in terms of these state-action pairs and a customer trajectory consists are a list of state-action pairs.

\begin{table}[t]
    \centering
    \resizebox{\columnwidth}{!}{ 
    \begin{tabular}{|c||c|}
    \hline
        \textbf{Feature Name} & \textbf{Possible Values} \\\hline\hline
        
        Primary location & \emph{web, mobile, teller, banker, ATM} \\\hline
        
        \makecell{Secondary location \\ (web and \\ mobile menus)} & \makecell{\emph{credit-card, credit-score, offers, rewards, operations, settings,} \\
        \emph{alerts-maintenance, contact-us, account-documents,} \\ 
        \emph{profile-maintenance}} \\\hline
        
        \makecell{Transitioning \\ actions} & \makecell{\emph{login, log-off, enter} \\
        \emph{enter-menu, exit, exit-menu}} \\\hline
        
        \makecell{Information- \\ gaining \\ actions} & \makecell{\emph{alerts-definition, alerts-history, atm-branches, balance,} \\ 
        \emph{benefits, demographic, documents, faq, help-call, help-email,} \\ 
        \emph{offers, credit-card-trans-history, credit-card-trans-summary,} \\
        \emph{limit-credit-card, credit-score-history,} \\ 
        \emph{messages, rewards-activity, rewards-use-points,} \\
        \emph{credit-score-summary, trans-history, trans-summary}} \\\hline
        
        \makecell{Modification \\ actions} & \makecell{\emph{demographic, password,  user-id}, \\ 
        \emph{limit-credit-card, alerts-definition}} \\\hline
        
    \end{tabular}
    }
    \caption{List of handcrafted features.}
    \label{tab:handcrafted_features}
\end{table}



\begin{figure*}
  \centering
    \includegraphics[width=0.54\textwidth]{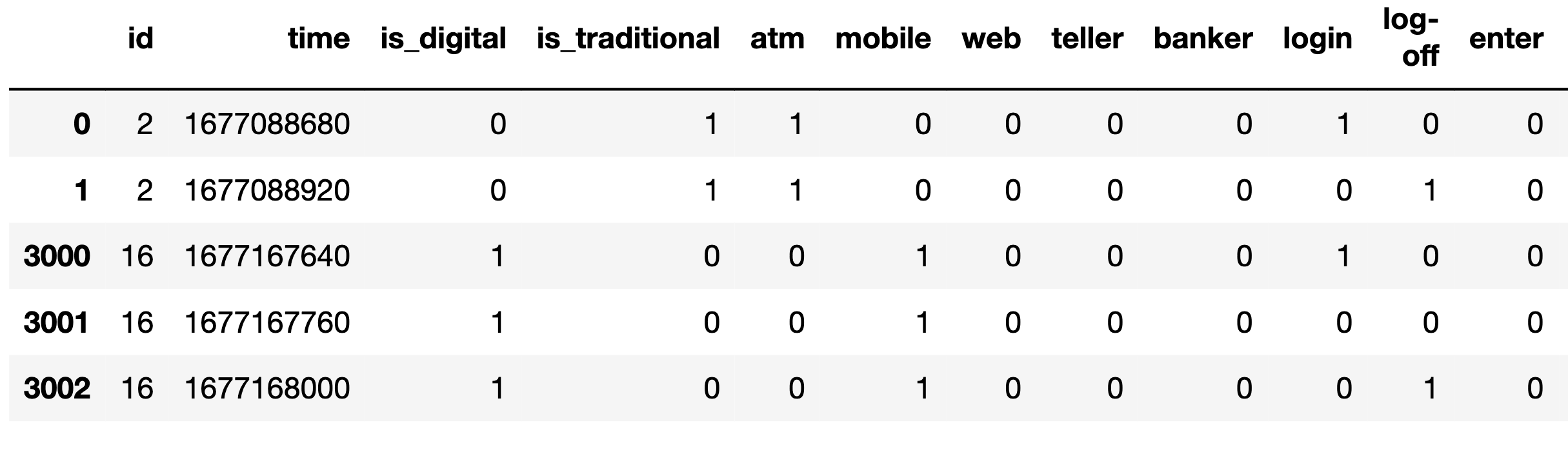}  
    \includegraphics[width=0.58\textwidth]{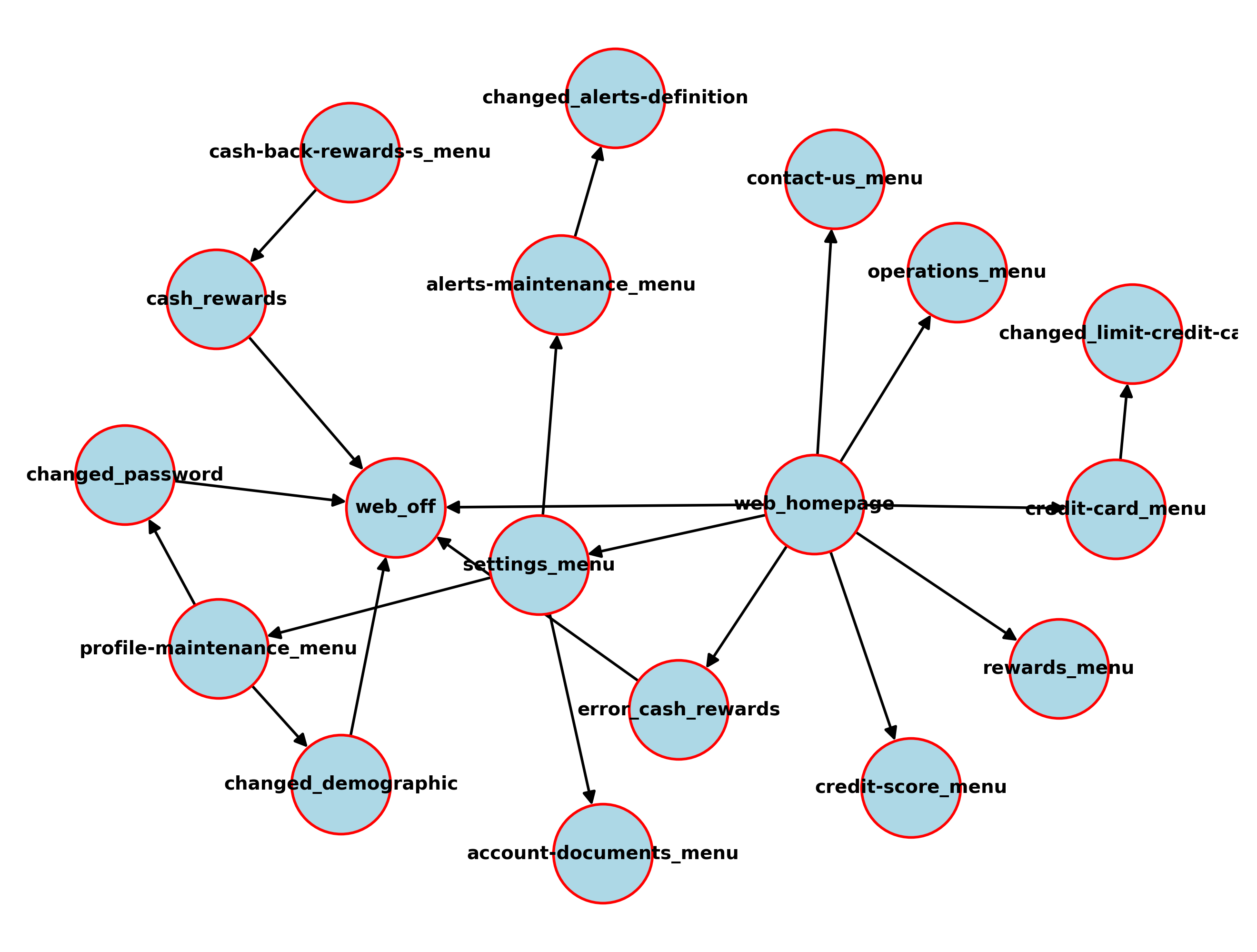}
  \caption{An example of bag-of-words and one-hot encoding representation, and a state-space graph representation.}
  \label{fig:data-repr}
\end{figure*}

\paragraph{Graph Representation:}

In addition to a bag-of-word representation \cite{harris1954distributional}, we also utilize a state graph embedding that will allow for better learning of each of our objectives.

In the graph embedding, nodes correspond to state features (i.e.,~primary and secondary location features). Nodes are connected by an edge if there exists a transition (i.e.,~an action that allows a customer to move between the two corresponding states). In addition to neighbor information, the node features of the graph are given as four binary indicators: the customer has visited this node in the past, the customer is currently at this node, the customer performed an information-gaining-action at this node, the customer performed a modification-action at this node. 
These indicators are refereed to respectively as $\langle$\emph{past-nodes}, \emph{ego-node}, \emph{info-gain}, \emph{modification}$\rangle$.

The intuition behind the advantage of this graph embedding is that it allows the model to have a less myopic understanding of the way in which actions affect the customer's location in each interface, which in turn allows the model to have a less myopic understanding of which actions will be available to the customer at the future time-steps as well as possible paths to nodes which may achieve the customer's goal.





\subsection{Predictive Models}

To capture the temporal patterns in customer interactions and learn the intricate relationships between actions and goals over time, we utilized the architecture of LSTM networks \cite{hochreiter1997long}. Specifically, we used a bag-of-words and one-hot encoding to represent the events as well as constructed a state graph to use as a GNN embedding in the LSTM model. The graph was constructed by examining all customer trajectories in the training dataset and defining nodes and edges as any state or transition action which appeared in more than 10 times. We found 10 to be the best frequency threshold when defining the graph as this value provided an effective balance between including irrelevant states and edges (which the model would then need to learn to ignore) and excluding important states and edges that will increase the myopicness of model predictions. Figure \ref{fig:data-repr} shows an example of bag-of-words representation and state-space graph representation, respectively, generated from the datasets.

\paragraph{Predictive Objective:} Our primary focus will be on the following predictive tasks:
\begin{itemize}
    \item \emph{Goal Prediction:} Given $n_H$ historical actions, what goal does the customer have?
    \item \emph{Type Prediction:} Given $n_H$ historical actions, what is the customer's type?
     \item \emph{Trajectory Prediction:} Given $n_H$ historical actions, what are the customer's next $n_F$ future actions?
\end{itemize}

\section{Experimental Results}

In this section, we present the experimental results of our two proposed models -- LSTM and GNN+LSTM. Our experimental setup consisted of the semi-synthetic dataset with 12 thousand customer interactions, split into 70\% training, 15\% validation, and 15\% test sets. Each data point in our dataset represents a sequence of a customer's previous 20 events. The LSTM and GNN+LSTM models were trained using the Adam optimizer \cite{kingma2014adam}, with a learning rate of 0.01, over 5000 epochs with early stopping based on the validation set performance. The primary metrics used for comparison were prediction accuracy for customer goals, agent types, and future events.

\smallskip
\noindent \textbf{Goal Prediction:} Table~\ref{tab:goal_pred} shows the accuracy for each approach when predicting the goal of the customer. We see that both the LSTM and GNN+LSTM models are capable of accurately predicting customer goals, however neither method achieves greater than 80\% accuracy. 
With that said, graph embedding does offer a significant improvement to model efficacy when compared to the bag-of-words embedding.

\begin{table}[t!]
    \centering \small
    \begin{tabular}{|c||c|c|}
    \hline
         \multirow{2}{*}{Model} & \multicolumn{2}{c|}{Accuracy} \\
         & Check Info & Change Info \\\hline\hline
         LSTM     & 71\% &  68\%\\\hline
         GNN+LSTM & 77\% &  75\%\\\hline
    \end{tabular}
    \caption{ Accuracy for each approach when predicting agent type. 
    LSTM corresponds to the LSTM approach with a bag-of-words embedding and GNN+LSTM corresponds to the LSTM approach with the graph representation and GNN encoding layer
    .
    For each prediction, the model views the customer's most recent 20 events.}
    \label{tab:goal_pred}
\end{table}

\smallskip
\noindent \textbf{Type Prediction:} In Table~\ref{tab:type_pred}, we see the accuracy when predicting customer type.\footnote{The ground truths for each customer type (income, fail behavior, digital behavior) were stated in the dataset.}
Both LSTM and GNN+LSTM  have high efficacy (roughly 90\% or more) when predicting both the agents failure rate type and preferred interface. 
In the case of preferred interface, customers tend to exclusively use their preferred interface, meaning that this predictive task becomes  significantly easier once observing historical events from each customer.

\smallskip
\noindent \textbf{Trajectory Prediction:} In Table~\ref{tab:traj_pred}, we see the accuracy when predicting the future events of customers.
While the accuracy on this task are notably lower than the other tasks, the space of possible predictions (i.e., the space of possible events) is far greater in trajectory prediction than in those other tasks. 
As is expected we see that as the model is required to forecast customer events farther into the future, its predictive efficacy decreases. 
However, we again see a large improvement in predictive efficacy when using the graph embedding over the bag-of-word embedding.

\begin{table}[t!]
    \centering \small
    \begin{tabular}{|c||c|c|c|}
    \hline
         \multirow{2}{*}{Model} & \multicolumn{3}{c|}{Accuracy} \\
         & Income & Fail Behavior & Digital Behavior \\\hline\hline
         LSTM     & 63\% & 89\% & 97\%\\\hline
         GNN+LSTM & 70\% & 90\% & 97\%\\\hline
    \end{tabular}
    \caption{Accuracy for each approach when predicting customer type.
    For each prediction, the model views the customer's most recent 20 events.}
    \label{tab:type_pred}
\end{table}

\begin{table}[t!]
    \centering \small
    \begin{tabular}{|c||c|c|c|}
    \hline
         \multirow{2}{*}{Model} & \multicolumn{3}{c|}{Accuracy} \\    
         & length 1 & length 5 & length 15 \\\hline\hline
         LSTM     & 52\% & 40\% & 33\%\\\hline
         GNN+LSTM & 67\% & 62\% & 49\% \\\hline
    \end{tabular}
    \caption{Accuracy for approach when predicting the next~1, 5, and 15 customer events.
    For each prediction, the model views the customer's most recent 20 events.}
    \label{tab:traj_pred}
\end{table}

\begin{figure}[t]
    \centering
    \includegraphics[width=\columnwidth]{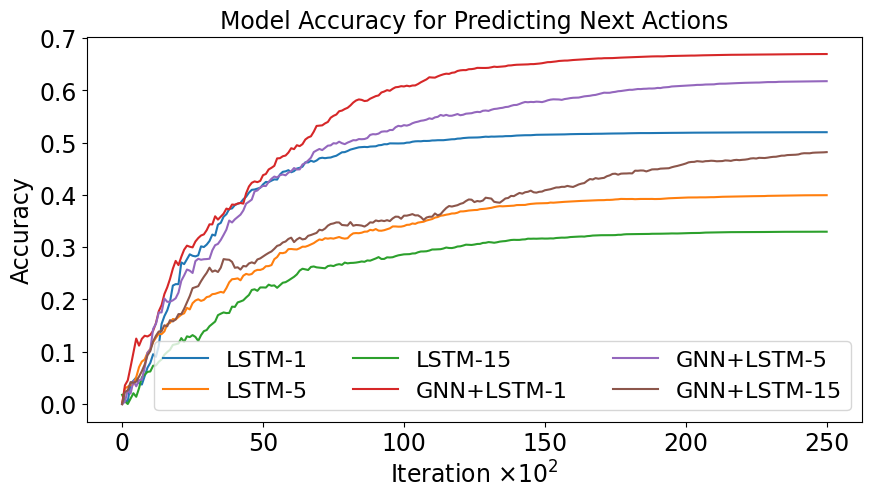}
    \caption{Loss function of LSTM and GNN+LSTM approaches when predicting the next 1, 5, and 15 actions of the customer when viewing the customer's last 20 actions.} 
    \label{fig:losses}
\end{figure}

When examining the bag-of-words approach with an LSTM model and the graph approach with a combination of a GNN and LSTM, we observe consistent and significant improvement with the graph embedding.
This is due in part to the non-myopic nature of the graph embedding in that it explicitly encodes all possible paths customers can take through each of the interfaces; as such the model to reason more efficiently about the ways in which current, or past events, influence future events. This is also visible in Figure \ref{fig:losses}, where we plot the loss functions of the LSTM and GNN+LSTM models.

\section{Conclusions}

We have provided a pipeline for predicting customer behavior, type, and objective, from observations of customers interacting with multiple bank interfaces (web, mobile, teller, ATM).
This pipeline is comprised of both a feature extraction procedure which takes reordered customer ``events" and builds a state-action graph as well as where states and actions are the result of hand-crafted features (Table~\ref{tab:handcrafted_features}). 
Second we use a combined GNN and LSTM architecture to make use of both the temporal and structural nature of customer interactions. 
We found this approach to be effective at each of the three predictive tasks, and to have consistent improvement over both baselines.

\paragraph{Future Work:} Currently we are exploring extensions of our graph based approach to not only predict customer behavior, but also to modify customer behavior. 
For example, the bank may desire to migrate customers towards digital interfaces, rather than in-person interfaces, i.e., modifying customer type.
To modify the behavior of customers which prefer in-person interfaces, we can first learn a state-action graph for those customers, which in turn allows us to surmise the perceived cost that those individuals have for using digital interfaces (unfamiliarity with mobile apps may increase and individual's perceived cost of taking actions within the bank's mobile app).
With these perceived costs,  the bank can then use targeted rewards (e.g., cash rewards for using the mobile app) to incentivize customers to use the mobile app.
After using the mobile app several times, individuals would become more familiar with app and would thus have a lower perceived cost of using the mobile app compared to in-person interfaces. 
Banks could then use this reward shaping technique to help migrate customers to a desired interface with greater precision than simply offering blind rewards to all individuals.
Behavior prediction, which is the current focus of this paper, is an essential first-step in modifying behavior as the ability to efficiently shape behavior is directly tied to the ability to predict that behavior.

\section{Acknowledgements}

This work was funded in part by J.P. Morgan AI Research. This paper was prepared for informational purposes by the Artificial Intelligence Research group of JPMorgan Chase \& Co and its affiliates (``J.P. Morgan''), and is not a product of the Research Department of J.P. Morgan. J.P. Morgan makes no representation and warranty whatsoever and disclaims all liability, for the completeness, accuracy or reliability of the information contained herein. This document is not intended as investment research or investment advice, or a recommendation, offer or solicitation for the purchase or sale of any security, financial instrument, financial product or service, or to be used in any way for evaluating the merits of participating in any transaction, and shall not constitute a solicitation under any jurisdiction or to any person, if such solicitation under such jurisdiction or to such person would be unlawful.

\bibliography{aaai23}

\end{document}